\documentclass[prl,nobalancelastpage,twocolumn,superscriptaddress,nofootinbib]{revtex4}

\usepackage{graphicx}
\usepackage{amsmath}
\usepackage{amssymb}
\usepackage[english]{babel}
\usepackage{color}

\newcommand{\Caltech}{Division of Physics, Mathematics and Astronomy, California Institute of Technology, Pasadena, CA 91125, USA}
\newcommand{\Harvard}{Department of Physics, Harvard University, Cambridge, MA 02138,
USA}
\newcommand{\mitaddress}{Department of Physics and Research Laboratory of Electronics, Massachusetts Institute of Technology,
Cambridge, MA 02139, USA}

\begin{document}
\title{Cold Matter Assembled Atom-by-Atom }
\author{Manuel Endres}\thanks{These authors contributed equally to this work.}
\address{\Harvard}
\address{\Caltech}
\author{Hannes Bernien}\thanks{These authors contributed equally to this work.}
\address{\Harvard}
\author{Alexander Keesling}\thanks{These authors contributed equally to this work.}
\address{\Harvard}
\author{Harry Levine}\thanks{These authors contributed equally to this work.}
\address{\Harvard}

\author{Eric R. Anschuetz}
\address{\Harvard}
\author{Alexandre Krajenbrink}\thanks{Current address:  CNRS-Laboratoire de Physique Th\'eorique de l'Ecole Normale Sup\'erieure, 24 rue L'homond, 75231 Paris Cedex, France}
\address{\Harvard}
\author{Crystal Senko}
\address{\Harvard}
\author{Vladan Vuletic}
\address{\mitaddress}
\author{Markus Greiner}
\address{\Harvard}
\author{Mikhail D. Lukin}
\address{\Harvard}
\begin{abstract}
The realization of large-scale fully controllable quantum systems is an exciting frontier in modern physical science. We use atom-by-atom assembly to implement a novel platform for the deterministic preparation of regular arrays of individually controlled cold atoms. In our approach, a measurement and feedback procedure eliminates the entropy associated with probabilistic trap occupation and results in defect-free arrays of over $50$ atoms in less than $400\,\text{ms}$. The technique is based on fast, real-time control of $100$ optical tweezers, which we use to arrange atoms in desired geometric patterns and to maintain these configurations by replacing lost atoms with surplus atoms from a reservoir. This  bottom-up approach enables controlled engineering of scalable many-body systems for quantum information processing, quantum simulations, and precision measurements.
\end{abstract}

\maketitle
The detection and manipulation of individual quantum particles, such as atoms or photons, is now routinely performed in many quantum physics experiments~\cite{Haroche:2012, Wineland:2012}; however, retaining the same control in large-scale systems remains an outstanding challenge. For example, major efforts are currently aimed at scaling up ion-trap and superconducting platforms, where high-fidelity quantum computing operations have been demonstrated in registers consisting of several qubits~\cite{Monroe2013,Devoret2013}. In contrast, ultracold quantum gases composed of neutral atoms offer inherently large system sizes. However, arbitrary single atom control is highly demanding and its realization is further limited by the slow evaporative cooling process necessary to reach quantum degeneracy. Only in recent years has individual particle detection~\cite{Bakr2010, Sherson2010} and basic single-spin control~\cite{Weitenberg2011} been demonstrated in low entropy optical lattice systems.

This Report demonstrates a novel approach for rapidly creating scalable quantum matter with inherent single particle control via atom-by-atom assembly of large defect-free arrays of cold neutral atoms~\cite{Weiss2004,Vala2015}. 
\begin{figure}[b!]
	\includegraphics[width=0.47\textwidth]{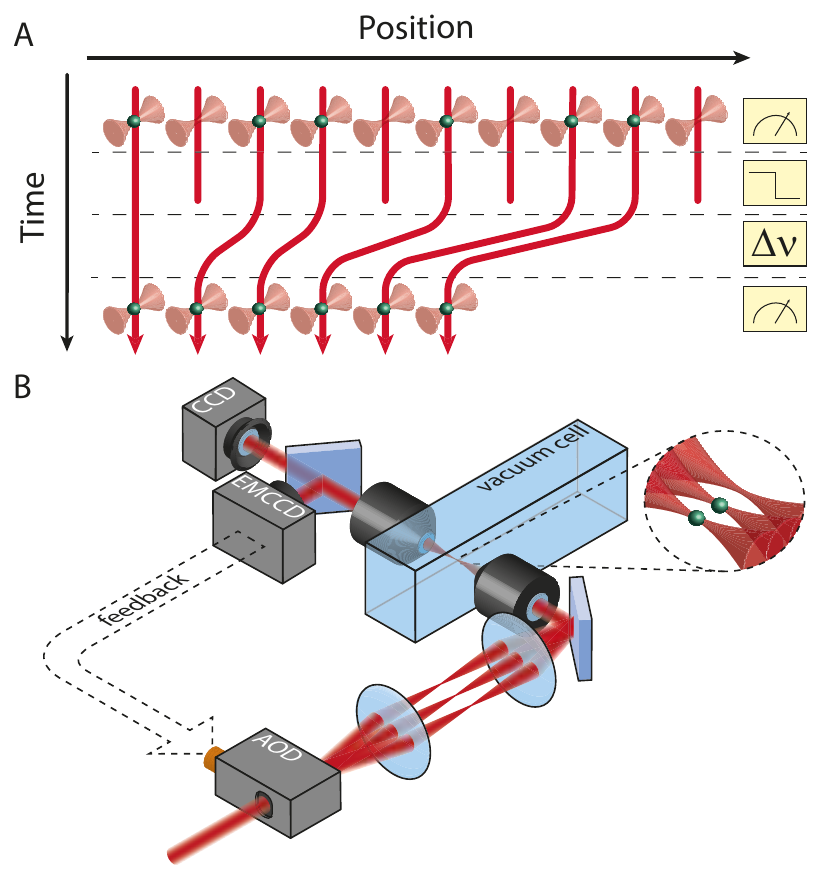}
	\caption{\textbf{Protocol  for creating  defect-free arrays.} (\textbf{A}) A first image identifies optical microtraps loaded with a single atom, and empty traps are turned off. The loaded traps are moved to fill in the empty sites and a second image verifies the success of the operation.
(\textbf{B}) The trap array is produced by an acousto-optic deflector (AOD) and imaged with a 1:1 telescope onto a $0.5$ NA microscope objective, which creates an array of tightly focused optical tweezers in a vacuum chamber. An identical microscope objective is aligned to image the same focal plane. A dichroic mirror allows us to view the trap light on a charge-coupled-device camera (CCD) while simultaneously detecting the atoms via fluorescence imaging on an electron-multiplied-CCD camera (EMCCD). The rearrangement protocol is realized through fast feedback onto the multi-tone radio-frequency (RF) field driving the AOD.}
\end{figure}
We use optical microtraps to directly extract individual atoms from a laser-cooled cloud~\cite{Frese2000, Schlosser:2001, Nelson:2007} and employ recently demonstrated trapping techniques~\cite{Schlosser2011, Piotrowicz2013, Nogrette2014,Kaufman2014, Kaufman2015} and single-atom position control~\cite{Miroshnychenko2006,Beugnon2007,Schlosser2012,Lee2016,Kim2016} to create desired atomic configurations. Central to our approach is the use of single-atom detection and real-time feedback ~\cite{Miroshnychenko2006,Lee2016,Kim2016} to eliminate the entropy associated with the probabilistic trap occupation~\cite{Schlosser:2001} (currently limited to ninety percent even with advanced loading techniques~\cite{Gruenzweig2010,Lester2015,Fung2015}). Related to the fundamental concept of ``Maxwell's demon"~\cite{Weiss2004,Vala2015}, this method allows us to rapidly create large defect-free atom arrays and to maintain them for long periods of time, providing an excellent platform  for large-scale experiments based on techniques ranging from Rydberg-mediated interactions~\cite{Jacksch:2000,Saffman:2010,Weimer2010, Browaeys:2016, Saffman2016} to nanophotonic platforms~\cite{Thompson2013b, Goban2014} and~Hubbard~model~physics~\cite{Kaufman2015,Kaufman2014, Murmann2015}.

\begin{figure*}[h!t]
	\includegraphics[width=1\textwidth]{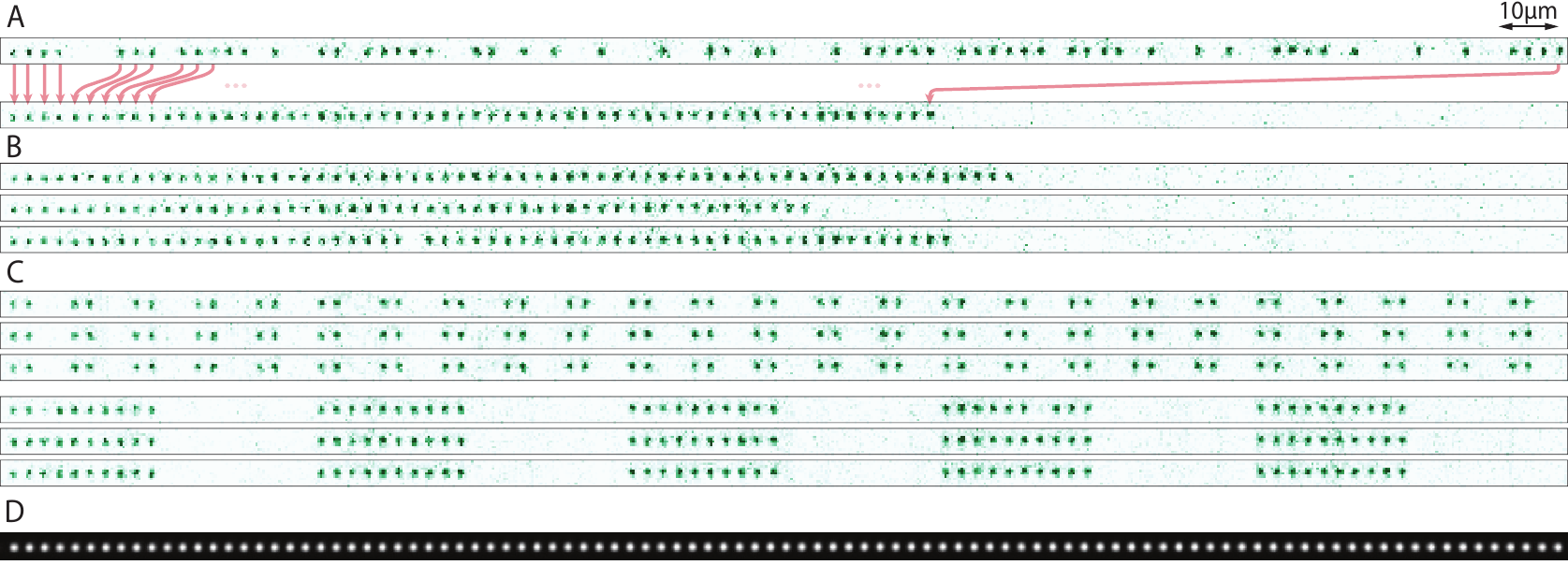}
	\caption{\textbf{Assembly of regular atom arrays.} (\textbf{A}) Single-shot, single-atom resolved fluorescence images recorded with the EMCCD before (top) and after (bottom) rearrangement. Defects are identified and the loaded traps are rearranged according to the protocol in Fig. 1, indicated by arrows for a few selected atoms. 
(\textbf{B}) Two instances of successfully rearranged arrays (first two pictures), and one instance (last picture) where a defect is visible after rearrangement.
(\textbf{C}) The final arrangement of atoms is flexible, and we generate, e.g., clusters of two (top) or ten (bottom) atoms. Non-periodic arrangements and adjustable lattice spacings are also possible.
(\textbf{D}) High-resolution CCD image of trap array. Our default configuration for loading atoms consists of an array of $100$ tweezers with a spacing of $0.49\,\text{MHz}$ between the RF-tones, corresponding to a real-space distance of $2.6\,\mu\text{m}$ between the focused beams~\cite{SOM}.
}
\end{figure*}

The experimental protocol is illustrated in Fig.~1A. An array of $100$ tightly focused optical tweezers is loaded from a laser-cooled cloud. The collisional blockade effect ensures that each individual tweezer is either empty or occupied by a single atom~\cite{Schlosser:2001}. A first high-resolution image yields single-atom resolved information about the trap occupation, which we use to identify empty traps and to switch them off. The remaining occupied traps are rearranged into a regular, defect-free array and we detect the final atom configuration with a second high-resolution image. 

Our implementation relies on fast, real-time control of the tweezer positions (see Fig.~1B), which we achieve by employing an acousto-optic deflector (AOD) that we drive with a multi-tone radio-frequency (RF) signal. 
\begin{figure}[h!t!]
	\includegraphics[width=0.49 \textwidth]{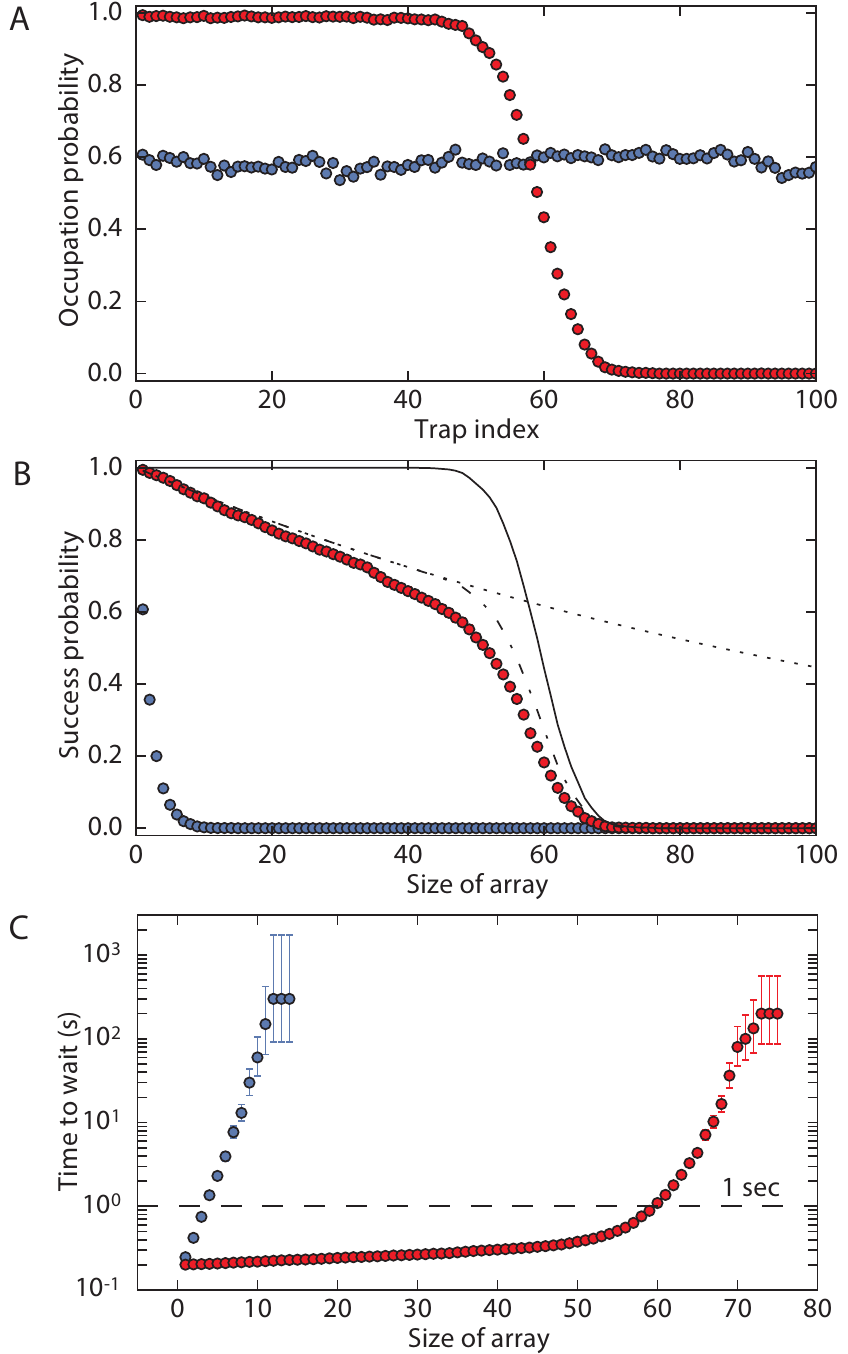}
	\caption{\textbf{Quantifying the rearrangement performance. }(\textbf{A}) The initial loading (blue circles) results in an occupation probability of $\approx 0.6$ for each trap in the array. After rearrangement (red circles), close to unity filling is reached on the left side of the array.
(\textbf{B}) In the initial image, the probability of finding a defect-free length-$N$ array (starting from the leftmost trap) falls off exponentially with $N$ (blue circles). Following the rearrangement of all loaded traps to form the largest possible array, we demonstrate strongly enhanced success probabilities at producing defect-free arrays (red circles). Theory curves show limits set by the total initial atom number (solid line), the background limited lifetime of $\tau=6.2\,\text{s}$ (dashed line) and the product of both (dashed dotted line)~\cite{SOM}.
(\textbf{C}) We show the expected amount of time to wait on average to produce a defect-free array of a given size taking into account the experimental cycle time of $200\,\text{ms}$ ($150\,\text{ms}$ without rearrangement). Without rearrangement, the wait time grows exponentially (blue circles). Employing the rearrangement procedure, we can produce arrays of length $50$ in less than $400\,\text{ms}$ (red circles). All error bars denote $68\%$ confidence intervals, which are smaller than the marker size in (A) and (B).
}
\end{figure}
This generates an array of deflected beams, each controlled by its own RF-tone~\cite{Kaufman2015,Kaufman2014}. The resulting beam array is then focused into our vacuum chamber and forms an array of optical tweezers, each with a Gaussian waist of $\approx 900\,\text{nm}$, a wavelength of $809\,\text{nm}$, and a trap depth of $U/k_B\approx 0.9\, \text{mK}$ ($k_B$, Boltzmann constant) that is homogeneous across the array within $2\%$~\cite{SOM}.

The tweezer array is loaded from a laser-cooled cloud of Rubidium-$87$ atoms in a magneto-optical trap (MOT). After the loading procedure, we let the MOT cloud disperse and we detect the occupation of the tweezers with fluorescence imaging. Fast, single-shot, single-atom resolved detection with $20\,\text{ms}$ exposure is enabled by a sub-Doppler laser-cooling configuration that remains active during the remainder of the sequence~\cite{SOM} (see Figs. 2A-2C). Our fluorescence count statistics show that individual traps are either empty or occupied by a single atom~\cite{Schlosser:2001,SOM}, and we find  probabilistically filled arrays with an average single atom loading probability of $p\approx 0.6$ (see Figs. 2A and 3A).

\begin{figure*}[h!t!]
	\includegraphics[width=1\textwidth]{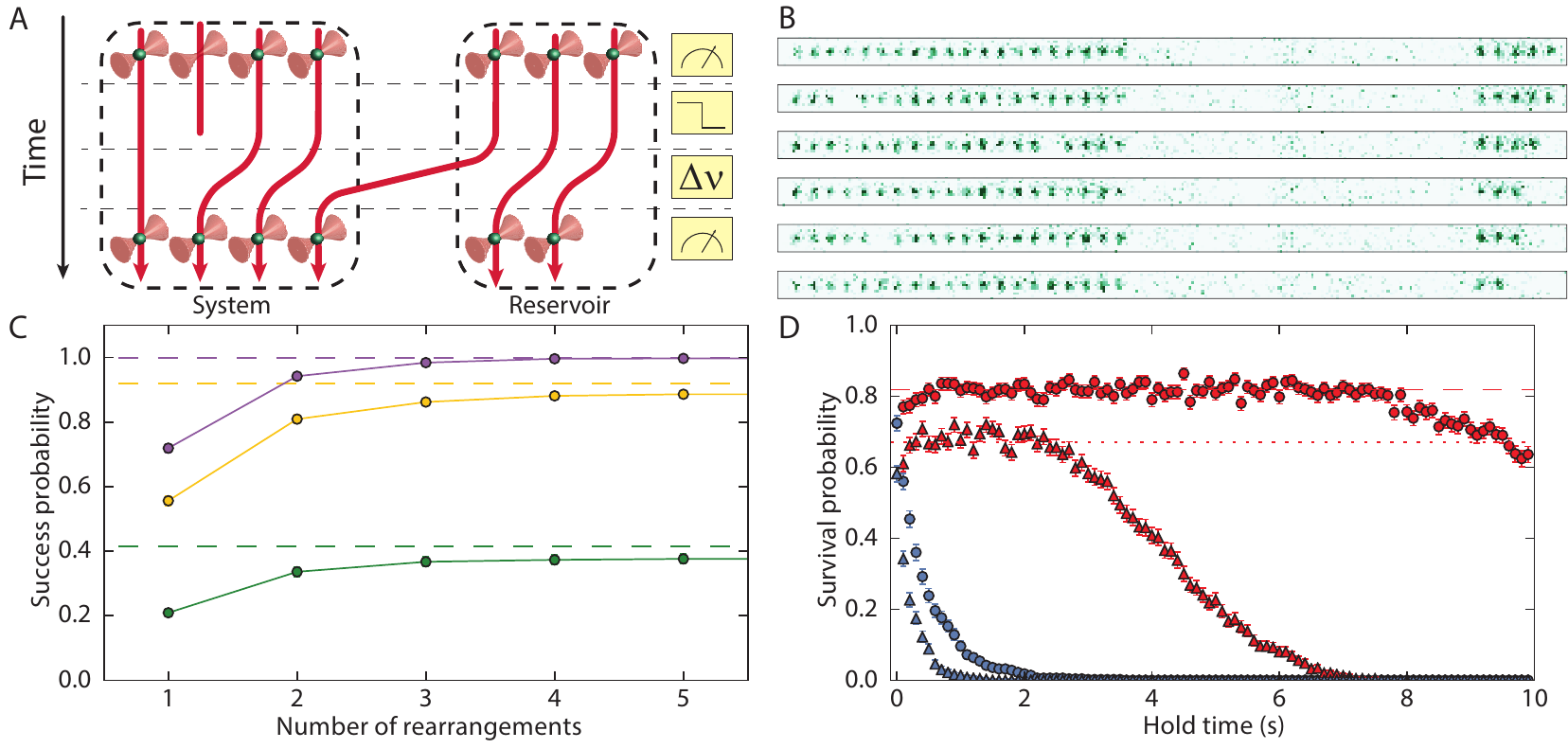}
	\caption{\textbf{Creating and maintaining regular arrays using an atomic reservoir.} (\textbf{A}) For a given target array size, surplus atoms are kept in a reservoir and used for repetitive reloading of the array.
(\textbf{B}) These images show a 20 atom target array with a reservoir of atoms. In this set of images, defects occasionally develop in the target array and are replaced by atoms in the reservoir. The reservoir depletes as it is used to fill in defects.
(\textbf{C}) By performing repeated rearrangements (once every $50\,\text{ms}$) the probability to successfully produce a defect-free array in any of these attempts increases and approaches the limit set by the number of initially loaded atoms (dashed lines). We show data for targeting 40 (purple), 50 (yellow), and 60 (green) atom arrays. Solid lines are guides to the eye.
(\textbf{D}) Probing for defects and filling them once every $100\,\text{ms}$ from the reservoir extends the lifetime of a defect-free array. We show data for the success probability of maintaining arrays of 20 (circles) and 40 (triangles) atoms with (red) and without (blue) replenishing atoms from the reservoir. With replenishing, the probability to maintain a defect-free array remains at a fixed plateau for as long as we have surplus atoms in the reservoir. The initial plateau value is set by the probability that no atoms in the array are lost in $100 \,\text{ms}$ (calculated value for $10\,\text{s}$ single atom lifetime shown as the dotted line). All error bars denote $68\%$ confidence intervals, which are smaller than the marker size in (C).
}
\end{figure*}

The central part of our scheme involves the rearrangement procedure for assembling defect-free arrays~\cite{SOM} (see Fig. 1A). In the first step,  unoccupied traps are switched off by setting the corresponding RF-amplitudes to zero. In a second step, all occupied tweezers are moved to the left until they stack up with the original spacing of $2.6\,\mu\text{m}$. This movement is generated by sweeping the RF-tones to change the deflection angles of the AOD and lasts $3\,\text{ms}$~\cite{SOM}. Finally, we detect the resulting atom configuration with a second high-resolution image. These steps implement a reduction of entropy via measurement and feedback. The effect is immediately visible in the images shown in Figs.~2A and 2B. While the initial filling of our array is probabilistic, the rearranged configurations show highly ordered atom arrays. Our approach also allows us to construct flexible atomic patterns as shown in Fig.~2C.

The rearrangement procedure creates defect-free arrays with high fidelity. This can be quantified by considering the improvement of single atom occupation probabilities (Fig.~3A) and the success probabilities, $p_N$, for creating defect-free arrays of length $N$ (Fig.~3B). The single atom occupation probability in the left-most forty traps increases from $\approx\!0.6$ before rearrangement to $0.988(3)$ after rearrangement, demonstrating our ability for high-fidelity single-atom preparation. Furthermore, the success probabilities for creating defect-free arrays show an exponential improvement. Prior to rearrangement, the probability of finding a defect-free array of length $N$ is exponentially suppressed with $p_N=p^N$ where $p \approx \! 0.6$ (blue circles, Fig.~3B). After rearrangement, we find success probabilities as high as $p_{30}=0.75(1)$ and $p_{50}=0.53(1)$ (red circles, Fig.~3B). 

The same exponential improvement is observed by considering the average wait time for producing defect-free arrays, given by $T/p_N$, where $T=200\,\text{ms}$ is the cycling time of our experiment~(see Fig.~3B). For example, we are able to generate defect-free arrays of $50$ atoms with an average wait time of less than $400\,\text{ms}$ (red circles, Fig.~3C). In contrast, the likelihood of finding defect-free arrays of this length without rearrangement is so small that the corresponding wait time would be on the order of several hundred years.

The success probabilities can be further enhanced through multiple repetitions of the rearrangement protocol. Figure 4 illustrates the procedure in which we target an atomic array of fixed length and create a reservoir from surplus atoms in a separate zone. After the initial arrangement of atoms into the target and reservoir zones, we periodically take images to identify defects in the target array and pull atoms from the reservoir to fill in these defects. This enhances our initial success probabilities at producing defect-free arrays within one MOT-loading cycle to nearly the ideal limit (see Fig.~4C).  

Finally, a similar procedure can be used for correcting errors associated with atomic loss. This becomes a significant limitation for large arrays since for a given lifetime of an individual atom in the trap $\tau$, the corresponding lifetime of the $N$ atom array scales as $\tau/N$. To counter this loss, we repeatedly detect the array occupation after longer time intervals and replenish lost atoms from the reservoir. As demonstrated in Fig.~4D, this procedure leads to exponentially enhanced lifetimes of our arrays.

These results demonstrate the ability to generate and control large, defect-free arrays at a fast repetition rate. The success probabilities are limited by two factors: the initial number of loaded atoms and losses during rearrangement. For example, the average total atom number in our array is $59 \pm 5$~\cite{SOM}, which results in the cutoff in the success probability in Fig.~3B starting from $N\approx 50$ (solid line). For shorter arrays, the fidelity is mostly limited by losses during rearrangement. These losses are  dominated by our finite vacuum-limited lifetime, which varies from $\tau\approx 6\,\text{s}$ to $\tau\approx 12\,\text{s}$ (depending on the setting of our atomic dispenser source), and are only minimally increased by the movement of the atoms~\cite{SOM}. The single atom occupation probability is correspondingly reduced by a factor $\exp(-t_r/\tau)$, where $t_r=50\,\text{ms}$ is the time for the whole rearrangement procedure~\cite{SOM}. This results in the success probabilities of creating length-$N$ arrays scaling as  $\exp(- t_r N/\tau)$, which dominates the slope for $N \lesssim 50$ in Fig.~3B (dashed line). Currently, we reach vacuum limited lifetimes only with sub-Doppler cooling applied throughout the sequence. However, the lifetime without cooling could be improved, for example, by using a different trapping laser and trapping wavelength~\cite{SOM}.

The size of the final arrays can be considerably increased by implementing a number of realistic experimental improvements. For example, the initial loading probability could be enhanced to $0.9$~\cite{Gruenzweig2010,Lester2015,Fung2015} and the vacuum limited lifetime could be improved to $\tau\approx 60\,\text{s}$ in an upgraded vacuum chamber. Increasing the number of traps in the current configuration is difficult due to the AOD bandwidth of $\approx 50\,\text{MHz}$ and strong parametric heating introduced when the frequency spacing of neighboring traps approaches $\approx 450\,\text{kHz}$~\cite{SOM}. 
However, implementing two-dimensional (2d) arrays could provide a path towards realizing thousands of traps, ultimately limited by the availability of laser power and the field of view of high-resolution objectives. Such 2d configurations could be realized by either directly using a 2d-AOD or by  creating a static 2d lattice of traps (using spatial light modulators~\cite{Nogrette2014} or optical lattices~\cite{Nelson:2007}) and sorting atoms with an independent AOD~\cite{SOM}. Realistic estimates for the rearrangement time $t_r$ in such 2d arrays indicate that the robust creation of defect-free arrays of hundreds of atoms is feasible~\cite{SOM}. Finally, the repetitive interrogation techniques, in combination with periodic reservoir reloading from a cold atom source (such as a MOT), could be used to maintain arrays indefinitely.

Remarkable recent advances have been made in building, exploring, and applying engineered quantum matter in systems ranging from ion traps~\cite{Monroe2013} and ultracold atoms~\cite{Bakr2010, Sherson2010,  Weitenberg2011} to  solid-state qubits such as superconducting qubits~\cite{Devoret2013} and color-centers~\cite{Awschalom2013}.  Atom-by-atom assembly of defect-free arrays forms a novel scalable platform with unique possibilities. It combines features that are typically associated with ion trapping experiments, such as single-qubit addressability~\cite{Xia2015, Wang2015} and fast cycling times, with the flexible optical trapping of neutral atoms in a scalable fashion. Furthermore, in contrast to solid-state platforms, such atomic arrays are formed of indistinguishable particles and are mostly decoupled from their environment. 

These features provide an excellent starting point for engineering interactions in multi-qubit experiments and for exploring novel applications. In particular, our approach can be used for the realization of logical qubits based on quantum registers consisting of up to several dozens of atomic spin qubits. The necessary entangling operations can be realized using atomic Rydberg states~\cite{Jacksch:2000,Saffman:2010, Browaeys:2016, Saffman2016}. The parallelism afforded by our flexible atom rearrangement  enables  efficient diagnostics of such Rydberg-mediated entanglement. For example, the use of atomic configurations demonstrated in Fig.~2C, supplemented by the repetitive reloading mechanism in Fig.~4, could allow fidelities to be optimized to their theoretical limit~\cite{Browaeys:2016, Saffman2016}. Similar techniques can be used for novel approaches to quantum simulations using both coherent coupling and engineered dissipation~\cite{Weimer2010,Browaeys:2016} and for creating large-scale maximally entangled quantum states for applications in precision measurements~\cite{Komar2016}. 

In addition, the realization of homogeneous arrays~\cite{SOM} should allow for simultaneous sideband cooling of atoms in optical tweezers~\cite{Kaufman2012,Thompson2013}. Ground state fractions in excess of $90\%$ have already been demonstrated, and can likely be improved via additional optical trapping along the longitudinal tweezer axes. This could enable bottom-up approaches to studying quantum many-body physics in Hubbard models~\cite{Kaufman2014, Kaufman2015, Murmann2015}, where atomic Mott insulators with fixed atom number and complex spin patterns could be directly assembled without the fluctuations and long cycling times associated with evaporative cooling. Moreover, these techniques could potentially be applied to create ultracold quantum matter composed of exotic atomic species or complex molecules~\cite{Barry2014,Hutzler2016} that are difficult to cool evaporatively. 
Finally, our atom arrays are ideally suited for integration with hybrid atom-nanophotonic platforms~\cite{Thompson2013b, Goban2014}. These can be employed to couple the atoms within a local multi-qubit register or for efficient communication between the registers using a modular quantum network architecture~\cite{Monroe2013}.

\bibliographystyle{h-physrev}
{
\renewcommand{\addcontentsline}[3]{}
\bibliography{bibliography}
}
-----------------------\\
\textbf{Acknowledgments} \\ We thank K.-K. Ni, N. Hutzler, A. Mazurenko and A. Kaufman for insightful discussion. This work was supported by NSF, CUA, NSSEFF and Harvard Quantum Optics Center. HB acknowledges support by a Rubicon Grant of the Netherlands Organization for Scientific Research (NWO).\\ 
-----------------------\\
\textbf{Supplementary Materials} \\ 
Materials and Methods\\
Figs. S1 to S5\\
Movies S1 to S3\\


{
\renewcommand{\addcontentsline}[3]{}
\section{Supplementary Material}
}

\setcounter{figure}{0}
\makeatletter 
\renewcommand{\thefigure}{S\@arabic\c@figure}
\makeatother

\tableofcontents

\section{1 Experimental Sequence}

\subsection{1.1 Trap loading}

We use an external-cavity diode laser at $\approx~$809~nm to seed a tapered amplifier (Moglabs, MOA002), which provides 1.8~W of output power. The resulting beam is coupled into a single-mode fiber, and passed through three laser clean-up filters (Semrock LL01-808). This results in a 4~mm beam with $\approx~$550~mW of power, which is split into an array of beams by an acousto-optic deflector (AOD) (AA Opto-Electronic model DTSX-400-800.850). These beams pass through a 1:1 telescope with 300~mm focal length and are then focused by a 0.5~NA microscope objective (Mitutoyo G Plan Apo 50X) into our vacuum chamber to form an array of optical tweezers (Fig.~1B of the main text). These tweezers have a waist of $\approx~$900~nm, and their centers are separated by 2.6$\,\mu$m. Each beam has $\approx~$1~mW of optical power, corresponding to a trap depth of $\approx~$0.9~mK for $^{87}$Rb atoms. 

The experimental sequence begins by laser cooling thermal $^{87}$Rb atoms in a magneto-optical trap (MOT) around the traps for 100 ms (Fig.~S1A). We use a gradient field of 9.7 G/cm, and three intersecting retroreflected beams that are 17 MHz red detuned of the free space $F=2 \to F'=3$ transition, overlapped with repumper beams resonant with the free space $F=1 \to F'=2$ transition. One of these beams is perpendicular to the optical table, and the other two are parallel to it (intersecting at an angle of $\approx~120^{\circ}$ due to the geometric restriction imposed by the high resolution objectives). All three beams have a diameter of $\approx~$1.5~cm, and carry $\approx~$1.5~mW of cooling light and $\approx~$0.4~mW of repumping light each. To reduce the necessary time to load the MOT, we shine UV light from a diode at 365~nm directly on the region of the glass cell within which the MOT is loaded \cite{Klempt2006_s}.

After 100~ms, the magnetic field gradient and the MOT beams are turned off to allow the MOT to dissipate over 28 ms. At the same time we turn on a set of probe beams which are 20 MHz red-detuned from the bare atom $F=2 \to F'=3$ transition. The probe beams have the same geometric configuration as the MOT beams, but they have $\approx~$50 times less power, and a diameter of $\approx~$1~mm, which largely reduces background light due to stray reflections during imaging. The probe beams further cool the atoms through polarization-gradient cooling, and are left on for the remainder of the sequence.

The result of this process is the probabilistic loading of atoms into the traps. For the data presented in Fig.~3 of the main text we loaded on average 59$\pm$5 atoms (Fig.~S1B).

\begin{figure}[t!!]
    \centering
    \includegraphics{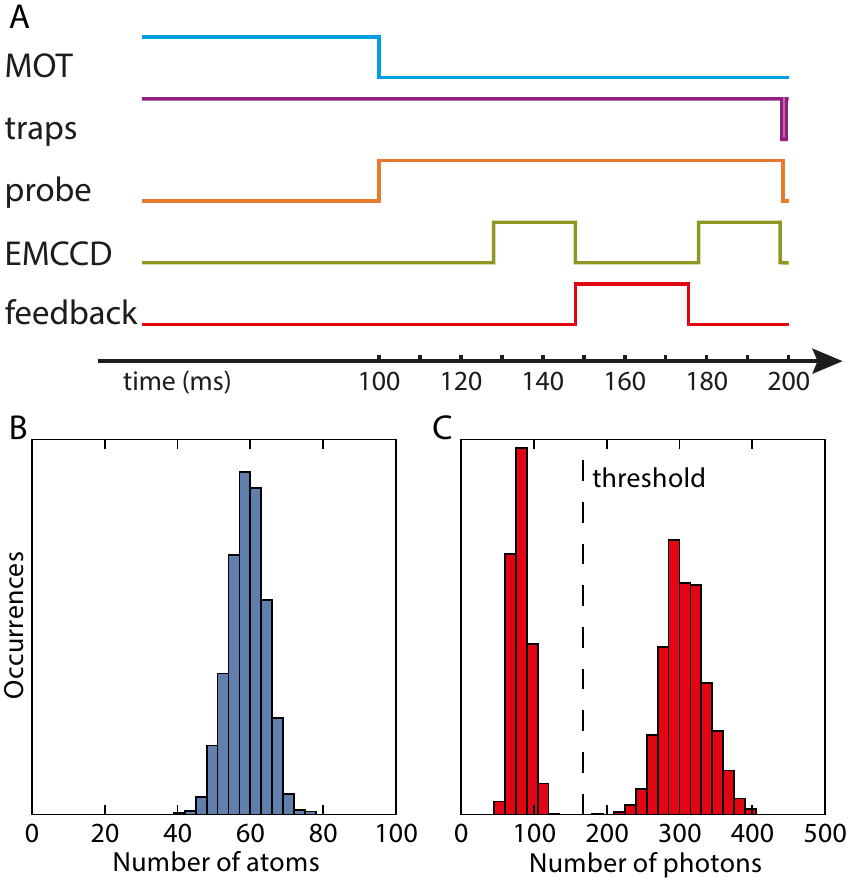}
    \caption{
    \label{fig:mot_loading}
    \textbf{Experimental sequence and atomic signal.}
    \textbf{A)} Pulse diagram of the experimental sequence. 
    \textbf{B)} Distribution of number of initially loaded atoms.
    \textbf{C)} Sample fluorescence count distribution for a single trap during 20~ms of exposure.}
\end{figure}

\subsection{1.2 Imaging}

The EMCCD camera (Andor iXON3) is triggered 128~ms after the beginning of the experimental sequence and acquires an image over the following 20~ms. Cooling light from the probe beams is scattered by the atoms and collected on the EMCCD, forming an image of the atoms in the array. Based on the collected photon statistics for each trap, we can set clear thresholds to determine the presence of an atom in a trap (see Fig.~S1C). Furthermore, the bimodal nature of the photon statistics is an indication that the traps are occupied by either 0 or 1 atom.

\subsection{1.3 Feedback}

Once the EMCCD has finished acquiring the signal, the image file is transferred over the following 10~ms to a computer which determines trap occupations, using pre-calibrated regions of interest and threshold counts for each trap, in sub-ms time. Using this information, the computer finds the correct pre-calculated waveform to displace each occupied trap during 3~ms from its initial position to its final position, and then adds up all of them into a multi-tone frequency-sweep waveform. This computation takes $\approx~$0.2~ms for each trap to be displaced. Once the waveform is ready, it is sent to the AOD to perform the trap displacement, and then it goes back to producing the original set of 100 traps. It is important to note that the rearrangement waveform is only calculated for loaded traps, which means that all empty traps are turned off for the duration of the rearrangement 
but are restored immediately afterwards, so that the trap array returns after rearrangement to its original configuration. After rearrangement, there is a $\approx~$7~ms buffer time before taking another 20~ms exposure image with the EMCCD. The entire process, consisting of image acquisition, transfer, analysis, waveform generation, rearrangement, and buffer, takes a total of 50~ms.

Currently, the profile of the frequency sweeps is calculated to be piece-wise quadratic in time, over a duration of 3~ms. For shorter transport times we observe an increase in the number of atoms lost during rearrangement. For the experiments reported in the main text, atom losses are only slightly increased compared to the expected lifetime in static traps (see Fig.~S2).
These losses depend on the distance that the atoms move, and it is possible that they could be reduced by changing the length or the profile of the frequency sweep to minimize acceleration and jerk during the trajectory. However, the fidelity of our rearrangement would not be significantly improved by minimizing these losses (see Fig. 3B of the main text).

\begin{figure}[t!!]
    \centering
    \includegraphics{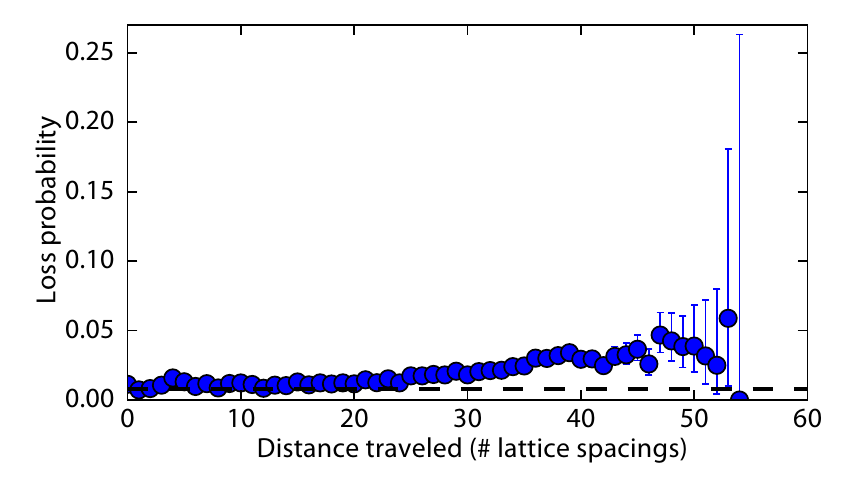}
    \caption{
    \label{fig:losses}
    \textbf{Rearrangement losses.}
    Losses as a function of total distance traveled by the atom during rearrangement for the dataset presented in Fig.~3 of the main text. The dashed line represents the expected loss for a stationary trap with 6.2 s lifetime.}
\end{figure}

\section{2 Experimental Methods}

\subsection{2.1 Generating 100 traps}

When driving the AOD with a single radio frequency (RF) tone, a portion of the input beam is deflected by an angle $\theta$ that depends on the frequency $\omega$ of the tone. By applying 100 different RF tones $\{\omega_1, \ldots, \omega_{100}\}$, we generate 100 beams with output angles $\{\theta_1, \ldots, \theta_{100}\}$, where $\theta_i = \theta(\omega_i)$. The waveform that we send into the AOD is initially calculated by a computer that samples the desired waveform in the time-domain with a sampling rate of 100 MHz. We stream these waveform samples to a Software Defined Radio (SDR) (Ettus Research, model USRP X310, daughterboard UBX~160) which performs digital-analog conversion, low-pass filtering, and subsequent analog IQ upconversion by a frequency of $\omega_{\text{up}}=74$~MHz, and outputs an analog waveform, which we then amplify and send to the AOD. The waveform that we calculate initially is given by:
\[
  \sum_{i=1}^{100} A_i e^{i \phi_i} e^{i (\omega_i - \omega_{\text{up}})t},
\]
with $A_i$ and $\phi_i$ being the real amplitude and phase, respectively, of the RF tone with frequency $\omega_i$. Since we generate all tones in the same waveform (relative to the same local oscillator inside the SDR), the tones in our waveform have well-defined phases $\{\phi_i\}$ relative to one another. Also, since all the frequencies we use are integer multiples of 1~kHz, we calculate a 1~ms waveform which is streamed on a loop without needing to continuously generate new samples.

\subsection{2.2 Effects of intermodulation}

The finite power bandwidth, along with other imperfections of our system (RF amplifier and AOD) generate a nonlinear response to the input signal. To the lowest order nonlinearity, the system acts as a mixer and generates new tones at the sum and difference of the input frequencies. For example, for two tones, $A_1 e^{i\phi_1} e^{i \omega_1 t}$,  $A_2 e^{i\phi_2} e^{i\omega_2 t}$, at the input, there will be a corresponding set of tones at the output:
\begin{eqnarray*}
  E^{-}_{1,2} =  A^{-}_{1,2} e^{i(\phi_1 - \phi_2)} e^{i(\omega_1 - \omega_2)t},
\end{eqnarray*}
\begin{eqnarray*}
  E^{+}_{1,2} =  A^{+}_{1,2} e^{i(\phi_1 + \phi_2)} e^{i(\omega_1 + \omega_2)t}
\end{eqnarray*}

These terms are far removed from the main set of desired tones $\{\omega_i\}$ and can in principle be filtered by frequency. However, they seed the next order of nonlinearity.

The next order of nonlinearity contains the mixing of these first order nonlinearities with the original tones. For example, with two tones we would now see a mixing of the $(\omega_1 - \omega_2)$ tone with the original $\omega_1$ tone to produce a sum tone at $(2 \omega_1 - \omega_2)$ and a difference tone at $\omega_2$. Similarly, the $(\omega_1 - \omega_2)$ tone would mix with the original $\omega_2$ tone to produce a tone at $(2\omega_2 - \omega_1)$ and $\omega_1$. If the phases of each input tone $\{\phi_i\}$ are not carefully selected, these intermodulations interfere destructively with the  original tones, as shown in Fig.~S3A.

\begin{figure*}[t!!]
    \centering
    \includegraphics{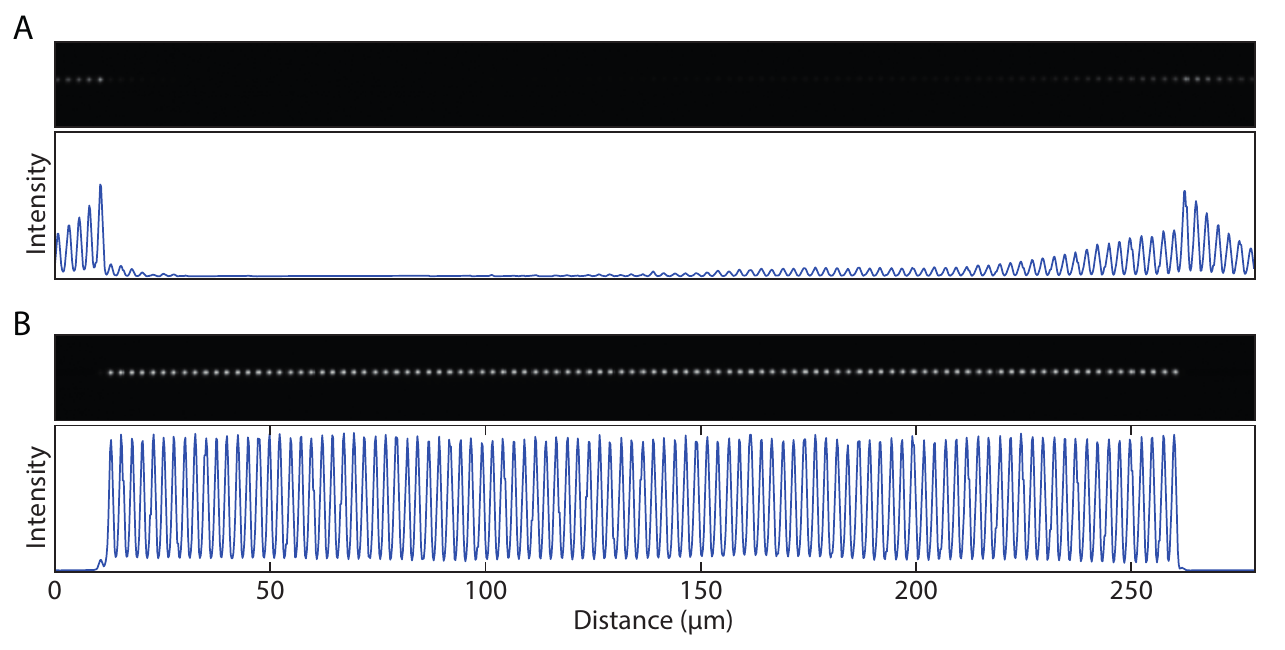}
    \caption{
    \label{fig:phases}
    \textbf{Generating uniform traps.} Nonlinearities inside the AOD and RF amplifier cause frequency mixing. (\textbf{A}) Setting identical phases for each input tone results in intermodulations that strongly interfere with the intended frequency tones and significantly distort the trap amplitudes. (\textbf{B}) By optimizing the phases $\{\phi_i\}$ and amplitudes $\{A_i\}$ of the RF tones we can reduce intermodulations and generate homogeneous traps.}
\end{figure*}

\subsection{2.3 Optimizing trap homogeneity}

We address the issue of intermodulations by adjusting the phases of the different RF tones to almost completely cancel out the nonlinearities. As a first step, we generate a computer simulation of 100 tones evenly spaced in frequency and with random phases, and equal amplitudes. For each pair of traps $\{i, j\}$, we calculate the difference tone $E^-_{i,j}$. By starting with random initial phases, the difference tones nearly completely destructively interfere with one another. We then optimize each phase, one at a time, to further reduce the sum of all difference tones $\sum E^-_{i,j}$. After this, we proceed to generate the waveform to be streamed onto the SDR. The starting amplitudes for each tone are selected such that they individually produce a single deflection carrying the correct amount of optical power. The frequencies span from 74.5 MHz to 123.01 MHz in steps of 0.49 MHz.

The next step in optimization consists of imaging the focused trap array on a CCD and performing 2D Gaussian fits. We use the amplitude of these fits to feed back on the amplitude of the individual tones. Once all the fitted amplitudes are approximately uniform, we continue to the last step of optimization.

Since we are interested in the uniformity of the traps at the positions of the atoms, we measure the AC Stark shifts induced by the traps, and use these values to feed back on the amplitude of each tone. In order to extract the AC Stark shift, we shine a single laser beam onto the trapped atoms for $10\,\mu\,$s, and measure the loss probability introduced by this ``pushout" beam as a function of detuning from the bare $F=2 \to F'=3$  resonance (Fig.~S4A inset). From the fits we extract the individual lightshift for each trap and use these values to perform feedback on the amplitudes. We repeat the procedure until the shifted resonances are uniform to within $\approx~ 2\%$ across the trap array (Fig.~S4A). At this point we have a set of optimal amplitudes $\{A^{opt}_{i}\}$ and phases $\{\phi^{opt}_{i}\}$ associated with the RF frequencies $\{\omega_{i}\}$ (Fig.~S3B). We interpolate between the values of $\{A^{opt}_{i}\}$ to define the optimal amplitude as a function of frequency $A^{opt}(\omega)$.
\subsection{2.4 Characterizing trap homogeneity}

\begin{figure}[t!!]
    \centering
    \includegraphics{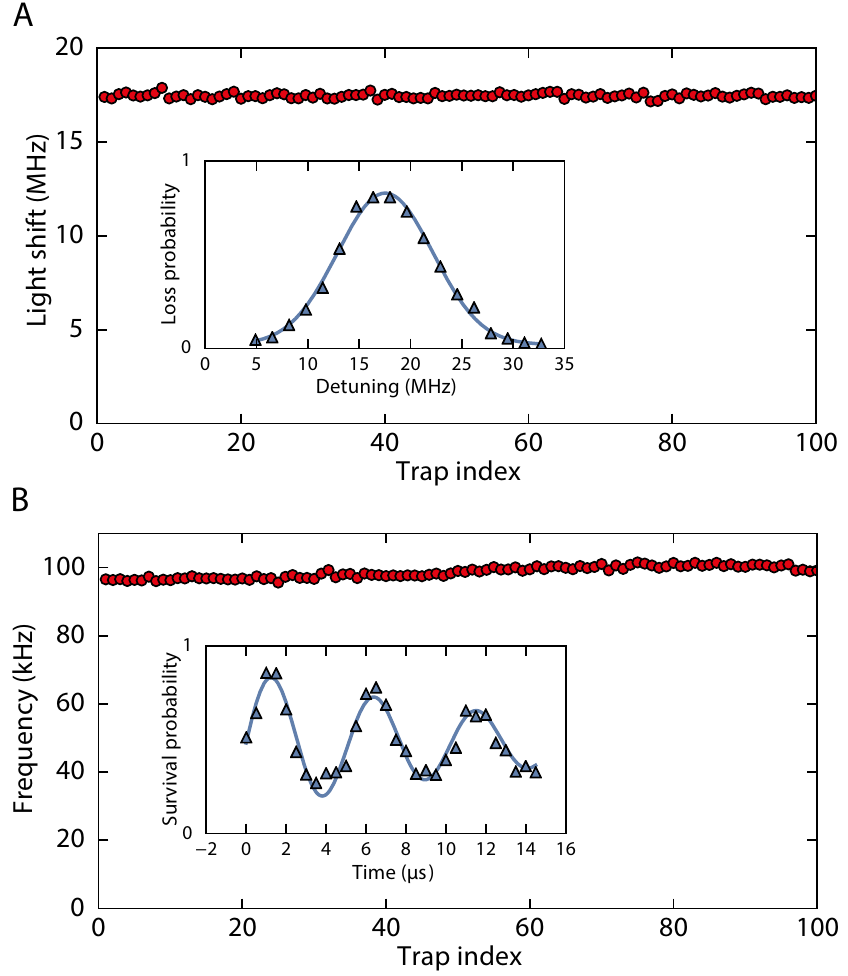}
    \caption{
    \label{fig:homogeneity}
\textbf{Characterization of trap and atom properties.} (\textbf{A}) The trapping light causes a lightshift on the atomic resonances that depends on the amplitude of the trap. The measured lightshifts across the array are used to optimize the RF amplitudes $\{A_i\}$. This process results in homogeneous traps with lightshifts that are uniform to within $\approx~$2\%. Inset shows the result of a “pushout” measurement (see text) on trap 26 that is used to determine the individual lightshift. (\textbf{B}) The trap frequencies are determined from a release and recapture measurement \cite{Sortais2007_s} (see inset for trap~26). The errors from the fits are smaller than the marker size for all the figures. }
\end{figure}

To characterize the homogeneity of the final trap configuration, we perform experiments to determine the AC Stark shift and trap frequency for each trap in the array (Fig.~S4).

As outlined in the previous section, the measurement of the AC Stark shift is used to equalize the traps. We find an average shift of 17.5 MHz with a standard deviation of 0.1 MHz across the whole array. 

The trap frequency (Fig.~S4B) is found through a release and recapture technique \cite{Sortais2007_s}. We obtain the radial frequency from a fit to the probability of retaining an atom as a function of hold time (Fig.~S4B~inset). We find the average trap frequency to be 98.7~kHz with a standard deviation of 1.7~kHz across the array. Combining these measurements with an independent determination of the waist, we estimate a trap depth of $\approx~$0.9~mK.

\subsection{2.5 Moving traps}

For most of the experimental sequence time, the traps are static. However, during short bursts we stream new waveforms to the AOD to rearrange them, and the atoms they hold. We move our traps by sweeping the frequencies $\{\omega_i(t)\}$ of the tones that correspond to the traps we wish to move, in a piecewise-quadratic fashion. This way, the atoms experience a constant acceleration $a$ for the first half of the trajectory, and $-a$ for the second half. During the sweep, we also continuously adjust the amplitude of the RF tone to match the optimized amplitude for its current frequency: $A_{i}(t) = A^{opt}(\omega_{i}(t))$. Further, by slightly adjusting the duration of each sweep, we enforce each trap to end with the optimal phase corresponding to its new position.

Using these parameters we pre-calculate all waveforms to sweep the frequency of a tone at any given starting frequency in our array to any given final frequency, over 3 ms. This amounts to $100^{2}$ precalculated trajectories.

\subsection{2.6 Lifetimes}
The lifetime for each trap is found by an exponential fit to the probability of retaining an atom as a function of time. Under optimal conditions, this results in an average lifetime for the traps in our array of 11.6~s with a standard deviation of 0.5~s across the array. However, this value depends on the background pressure inside the vacuum chamber and therefore depends on the current with which we drive our dispenser atom source. For the measurements presented in Fig.~3 of the main text, an average lifetime of 6.2~s was found from independent calibration measurements.

In these measurements, we apply continuous laser cooling throughout the hold time. We observe that without continuous cooling, the lifetime is reduced compared to these values. The retention as a function of time in this case does not follow a simple exponential decay (indicative of a heating process), and we define the time at which the retention probability crosses $1/e$ to be the lifetime. For the standard configuration in the main text (100 traps with 0.49~MHz spacing between neighboring frequencies), we find a lifetime of $\approx 0.4\, \text{s}$. While we have not characterized the source of the lifetime reduction in detail, we have carried out a number of measurements in different configurations to distinguish various effects:
\begin{itemize}
\item Generating a single trap by driving the AOD with a single frequency from a high-quality signal generator (Rhode \& Schwartz SMC100A), we find a lifetime of $\approx 2\,\text{s}$. This indicates that there are additional heating effects, such as photon scattering from trap light or intensity noise, that are independent of the use of the SDR or the fact that we drive the AOD with a large number of frequencies. Possible improvements include using further detuned trap light and improving our intensity stabilization. Furthermore, in a separate experiment~\cite{Thompson2013b_s}, we observed that using a Titanium-Sapphire laser instead of a TA significantly improved trap lifetimes even at the same trapping wavelength.
\item Generating a single trap by driving the AOD with a single frequency from the SDR, we find a lifetime of $\approx 1\,\text{s}$. This indicates that the RF-waveforms from the SDR could have additional phase or intensity noise. A possible source is the local oscillator used for IQ upconversion in the SDR, which could be replaced with a more phase-stable version.
\item We observe the same trap lifetime of $\approx 1\,\text{s}$ when driving the AOD with 70 frequencies at a spacing of $0.7$ MHz using the SDR. This indicates that, in principle, there is no lifetime reduction associated with driving the AOD with a large number of frequencies.
\item We observe a reduction of lifetime for frequency spacings smaller than $\approx 0.65 \,\text{MHz}$. For example, we found a lifetime of $\approx 0.4\, \text{s}$, when setting the frequency spacing to $0.49\, \text{MHz}$. (This lifetime is unchanged by increasing the number of traps from 70 to 100.) We interpret this effect to be the result of interference between neighboring traps. Due to the finite spatial overlap of the tweezer light, a time-dependent modulation occurs with a frequency given by the frequency spacing between neighboring traps. When bringing traps closer together, both the spatial overlap increases and the modulation frequency approaches the parametric heating resonance at $\approx 200 \text{kHz}$ given by twice the radial trapping frequency. 
\end{itemize}
We would like to stress that these effects only play a role after the rearrangement procedure is completed. During rearrangement, continuous laser cooling is a valid and powerful method to reduce heating effects. Additionally, the flexibility of our system makes it possible to load and continuously cool atoms in a set of closely spaced traps, and to rearrange the filled traps to an array with larger separations, at which point cooling can be turned off. This method of rearrangement takes advantage of the large number of atoms that can be initially loaded in our set of 100 traps separated by 0.49 MHz, while eliminating the effect of interference by setting a larger final frequency separation of, for example, 0.7 MHz. Furthermore, atoms could also be transferred into a fully ``static" trap array, such as an optical lattice, after rearrangement. 

However, even with optimal cooling parameters, heating effects cannot be always compensated. For example, we observed a significant reduction of lifetime for frequency spacings below $\approx 0.45~\text{MHz}$ even with continuous laser cooling. This sets a limit on the maximum number of traps that can be generated within the bandwitdh of our AOD, and therefore limits the final sizes of the atomic arrays.

\section{3 Prospects for extensions to 2D arrays}

In this section we will discuss possible extensions of our method to form uniformly filled two-dimensional (2d) arrays. We will describe two different rearrangement strategies and compare their performance given realistic parameters for loading efficiencies and lifetimes.

\subsection{3.1 Method 1: Row or column deletion and rearrangement}

Using a 2d AOD we could generate a 2d array of optical tweezers using two sets of RF tones, one set corresponding to rows and the other corresponding to columns. After loading atoms probabilistically into the array, it would be possible to eliminate each defect by turning off the RF frequency that generates either the row or the column containing the defect, and then transport all remaining rows and columns to form a defect-free uniform array.

\subsection{3.2 Method 2: Row-by-row rearrangement}

A different approach is to generate a static two-dimensional array of traps using techniques such as optical lattices or spatial-light modulators. In a first step, atoms would be probabilistically loaded into this static array. 
After loading, we could use an independent AOD to generate a linear array of traps, deeper than the ones forming the static array and overlapping precisely with one row of static traps. By rearranging the linear array, we could transfer the atoms to their final locations in the static configuration, where they would remain after turning off the traps used for transport. Doing this for each row would make it possible to fill an entire region of the static 2d array.

\begin{figure*}[t!!]
    \centering
    \includegraphics{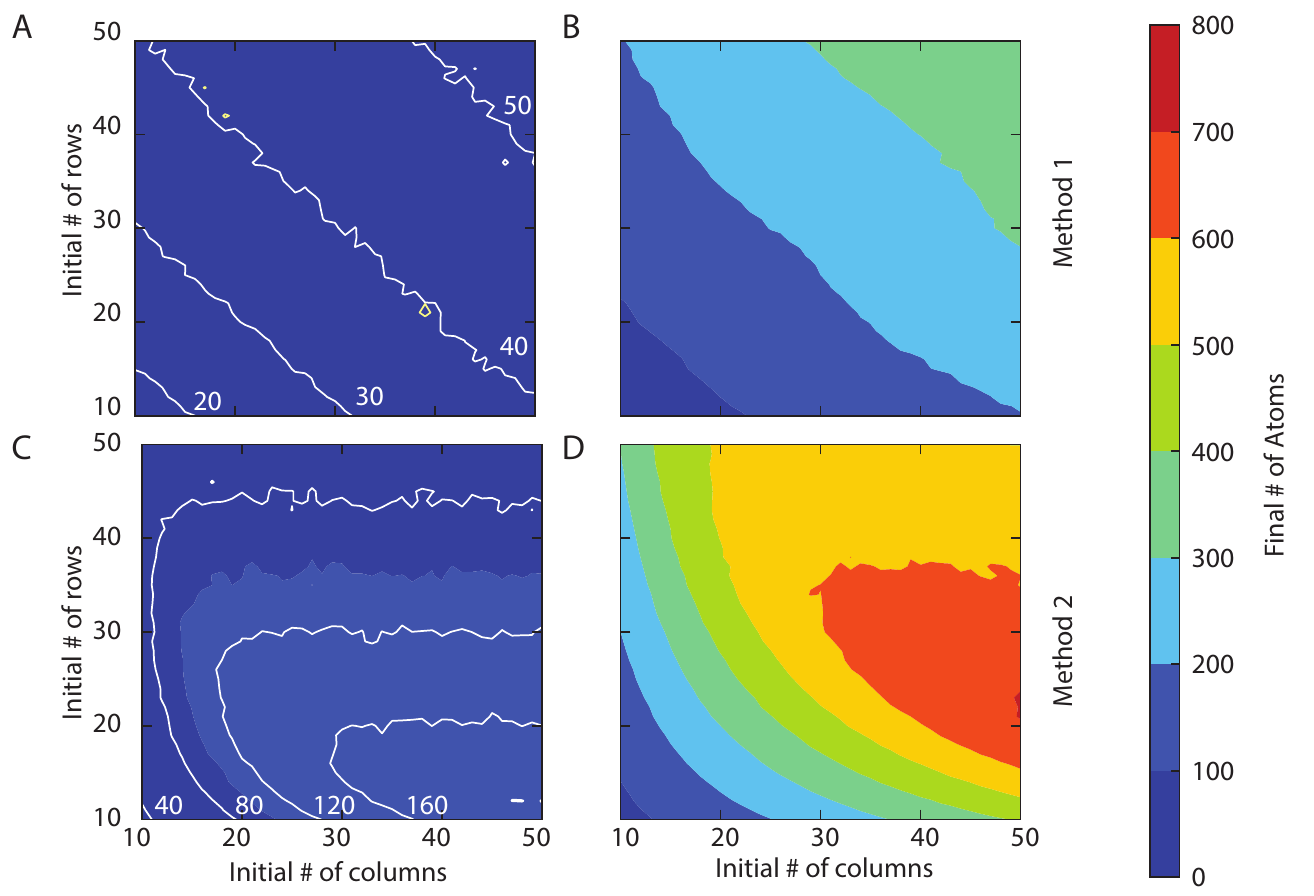}
    \caption{
    \label{fig:2d}
    \textbf{Projections for 2D system sizes.} Expected number of atoms in defect free rectangular arrays generated through the ``Row or column deletion and rearrangement method" with 0.6 loading efficiency and 10 s lifetime
    (\textbf{A}), and  with 0.9 loading efficiency and 60 s lifetime
    (\textbf{B}). Expected number of atoms in defect free rectangular arrays generated through the ``Row-by-row rearrangement method" with 0.6 loading efficiency and 10 s lifetime
	(\textbf{C}), and  with 0.9 loading efficiency and 60 s lifetime
    (\textbf{D}).}
\end{figure*}

\bibliographystyle{h-physrev}

\subsection{3.3 Expected performance and scalability}

The final size of the array will depend on the initial loading efficiency, while the probability to have a defect-free array after rearrangement will depend on the atom lifetimes in the traps, and the total feedback time. This time consists of several blocks: image taking, image transfer, waveform calculation, and trap movement. We can take an image in 20~ms. The time it takes to transfer the image from the camera to the computer takes a minimum of 9 ms and each row of atoms adds 0.8 ms to the transfer time. Analyzing the image to determine the location of the atoms and necessary frequency sweeps can be done in sub-ms time. Generating the waveform takes $\approx~$0.2~ms for each sweep necessary. This time scales as the final number of atoms ($O(N)$) for method~2, and with the final number of rows and columns ($O(N^{1/2})$) for method~1. Finally, the rearrangement itself takes 3~ms for each set of frequency sweeps: for method 1 there are two sweeps, and for method 2 the number of sweeps equals the number of rows.

Figure S5 shows the result of a Monte Carlo simulation for the expected number of atoms in a defect-free rectangular array using two sets of parameters, and both rearrangement methods described above. Given our current loading efficiency of $0.6$ and vacuum-limited lifetime of $\approx~$10~s, we can expect more than 160 atoms in the final defect-free configuration (Fig.~S5A,~C). However, if we were to upgrade our vacuum setup to increase the lifetime from 10~s to 60~s, and we increased the loading efficiency to 0.9 using currently available techniques \cite{Gruenzweig2010_s,Lester2015_s,Fung2015_s}, we could expect more than 600 atoms in defect-free configurations (Fig.~S5B,~D). These numbers were calculated by simulating a sequence that performed repeated rearrangement until no more defects appeared, and every point on the plot is the average of 500 simulations.

\section{4. Movies of the rearrangement procedures}
{\it(See ancillary files)}

VS1: Rearrangement procedure. This video shows consecutive pairs of atom
fluorescence images. Each pair consists of an initial image showing the
random loading into the array of traps and a subsequent image after the
loaded atoms have been rearranged to form a regular array.  The sequence
of images demonstrates the effectiveness of the procedure to reduce the
entropy associated with the random initial positions of the atoms by
ordering them in long, uninterrupted chains. The 1 Hz cycle rate in the video is slowed down by a factor of five relative to the 5 Hz experimental repetition rate.

VS2: Lifetime extension of 20 atom array. This video shows consecutive
fluorescence images of atoms being rearranged from an initial
probabilistic loading into an ordered array of 20 atoms and a reservoir
formed by surplus atoms. The length of the target array is kept constant
by transporting atoms from the reservoir to replace lost atoms in the
target array. This technique makes it possible to extend the average
lifetime of a defect-free 20 atom array to $\approx~$8 s, at which point the
reservoir typically has depleted. The 10 Hz frame rate of this video
accurately reflects the rate at which these images were acquired.

VS3: Lifetime extension of 40 atom array. This video shows consecutive
fluorescence images of atoms being rearranged from an initial
probabilistic loading into an ordered array of 40 atoms and a reservoir
formed by surplus atoms. The length of the target array is kept constant
by transporting atoms from the reservoir to replace lost atoms in the
target array. This technique makes it possible to extend the average
lifetime of a defect-free 40 atom array to $\approx~$2 s, at which point the
reservoir typically has depleted. The 10 Hz frame rate of this video
accurately reflects the rate at which these images were acquired.

\renewcommand{\addcontentsline}[3]{}

\end{document}